\documentclass[aps,prl,twocolumn,superscriptaddress,groupedaddress,showpacs]{revtex4-1}

\pdfoutput=1 

\usepackage{graphicx}  
\usepackage{dcolumn}   
\usepackage{bm}        
\usepackage{amssymb}   

\newcommand{\be}{\begin{equation}}
\newcommand{\ee}{\end{equation}}
\newcommand{\bea}{\begin{eqnarray}}
\newcommand{\eea}{\end{eqnarray}}

\usepackage{setspace}
\usepackage{amsmath}
\usepackage{amssymb}
\usepackage{fancybox}
\usepackage{xcolor}

\usepackage[protrusion=false,tracking=false,kerning=true,spacing=false,factor=1000,stretch=10,shrink=40]{microtype}
\usepackage[english]{babel}
\usepackage{amsmath,amsthm,amssymb}
\usepackage{amsfonts}
\usepackage{amsopn}
\usepackage{float}
\usepackage{setspace}
\usepackage{fancybox}
\usepackage{url}
\usepackage{appendix}
\usepackage{graphicx}
\usepackage{dcolumn}
\usepackage{bm}
\usepackage{color}
\usepackage{esint}

\usepackage{float}

\begin{document}
\author{Maciej Koch-Janusz}
\affiliation{Department of Condensed Matter Physics, Weizmann Institute of Science, Rehovot IL-76100, Israel}
\author{D. I. Khomskii}
\affiliation{II.Physikalisches Institut, Universitaet zu Koeln, Germany}
\author{Eran Sela}
\affiliation{Raymond and Beverly Sackler School of Physics and Astronomy, Tel-Aviv University, Tel Aviv, 69978, Israel}

\title{Two-dimensional Valence Bond Solid (AKLT) states from $t_{2g}$ electrons}

\begin{abstract}
Two-dimensional AKLT model on a honeycomb lattice has been shown to be a universal resource for quantum computation.
 In this valence bond solid, however, the spin interactions involve higher powers of the Heisenberg coupling $(\vec{S}_i \cdot \vec{S}_j)^n$, making these states seemingly unrealistic on bipartite lattices, where one expects a simple antiferromagnetic order. We show that those interactions
can be generated by orbital physics in multiorbital Mott insulators. We focus on $t_{2g}$ electrons on the honeycomb lattice and propose a physical realization of the spin-$3/2$ AKLT state. We find a phase transition from the AKLT to the Neel state on increasing Hund's rule coupling, which is confirmed by density matrix renormalization group (DMRG) simulations. An experimental signature of the AKLT state consists of protected, free spins-1/2 on lattice vacancies, which may be detected in the spin susceptibility.
\end{abstract}
\pacs{75.10.Nr,  75.10.Jm, 75.10.Hk}

\maketitle

\paragraph{Introduction \--}
Spin liquids  are ground states of magnetic systems which do not exhibit magnetic symmetry breaking down to zero temperature \cite{balentsNature}; in contrast to conventional ordered states like ferromagnets and antiferromagnets (AF), they host fractional excitations \cite{PhysRevB.39.11413}. A classical example is the spin-$1$ Heisenberg chain
 which, as predicted by Haldane \cite{Haldane1983464,PhysRevLett.50.1153} and observed in experiment \cite{PhysRevLett.65.3181}, does not order and exhibits fractionalized spin-$1/2$ excitations at its boundaries.

\begin{figure}[h]
\centering
\includegraphics*[width=8.5cm]{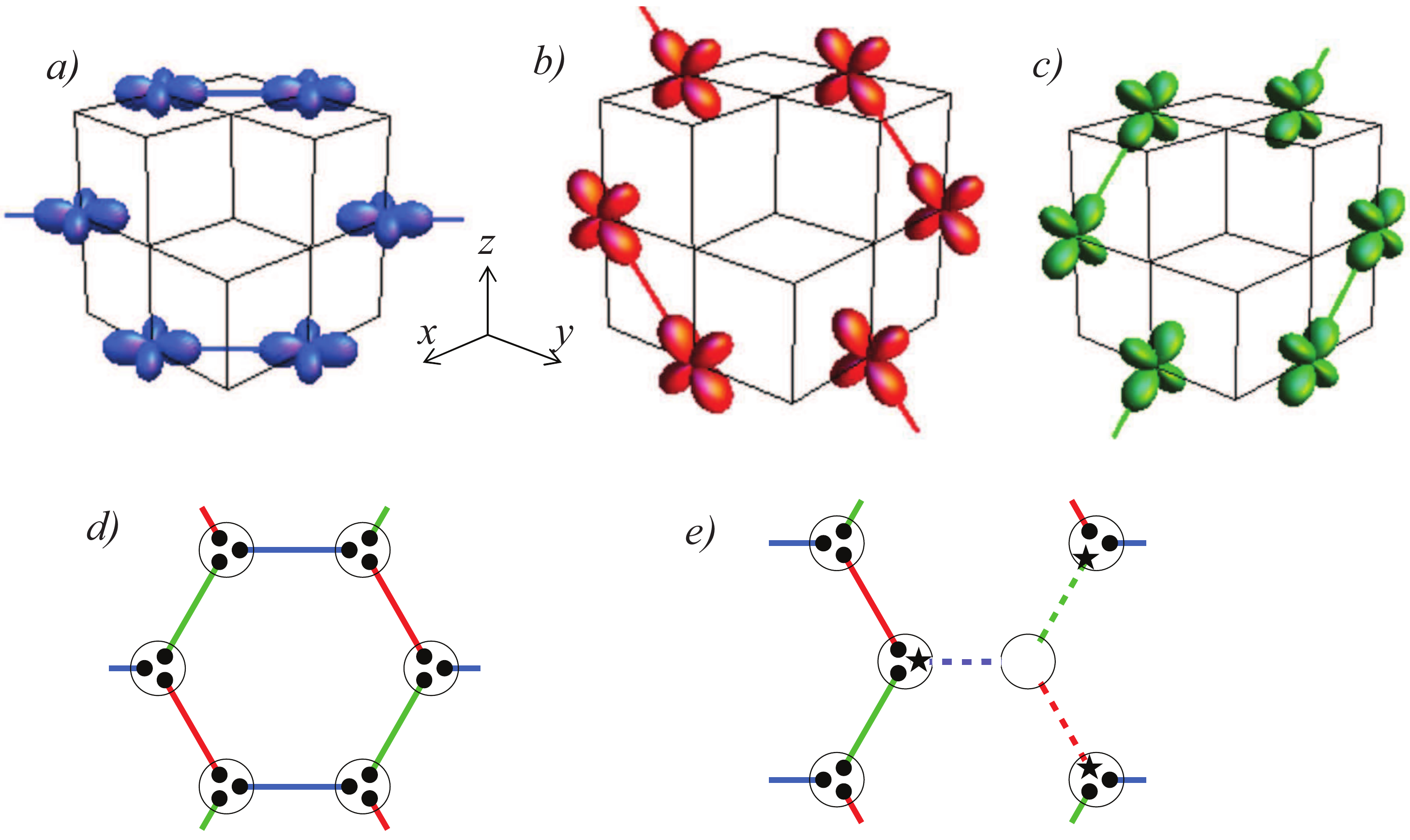}
\caption{(Color online) Nearest-neighbout $t_{2g}$ hoppings, corresponding to (a) $xy$, (b) $yz$, and (c): $zx$ orbitals, and (d) the resulting directional-hopping model, (e) free spins (denoted by stars) around lattice vacancy in the AKLT phase.}
\label{fig:t2g3plots}
\end{figure}

 In 1987, Affleck, Kennedy, Lieb, and Tasaki (AKLT) formulated a family of exactly solvable valence-bond-solid (VBS) models which (i) account for the Haldane phase in one dimension \cite{PhysRevLett.59.799}, but also (ii) generalize to higher dimensions  \cite{AKLTmath}. In these VBS states, the spin-$S$ at each site is partitioned into $2S$ spins-$1/2$, and the latter form a solid of singlet valence bonds, see Fig \ref{fig:t2g3plots}d for the $S=3/2$ case. As a result the spin-singlet ground state exhibits spin-$1/2$ excitations at edges or boundaries, whenever valence bonds are broken.
The model Hamiltonians which realize VBSs as exact ground states consist of a sum of operators projecting the total spin on neighbouring sites to the maximal possible value of $2S$. These projectors may be expanded in powers of $(\vec{S}_i \cdot \vec{S}_j)$; for the AKLT model of spins-3/2 on the honeycomb lattice which will be considered here it reads:
\be
\label{eq:AKLT32}
H_{AKLT}=\mathcal{J}\sum_{\langle i,j \rangle} [\vec{S}_i \cdot \vec{S}_{j} + c_2 (\vec{S}_i \cdot \vec{S}_{j})^2+ c_3 (\vec{S}_i \cdot \vec{S}_{j})^3],
\ee
with $c_2=
\frac{116}{243}$ and $c_3=
\frac{16}{243}$. The multi-quadratic interactions are typically not found in real materials e.g. insulators; usually the dominant spin-spin interaction is generated by superexchange, leading to AF Heisenberg interactions. Consequently, except for the 1D case in which the absence of magnetic ordering at T=0 is guaranteed by strong quantum fluctuations, the ground states on bipartite lattices break the spin $SU(2)$ symmetry and order antiferromagnetically \cite{AKLTmath}. As a result, a physical realization of the 2D AKLT phase in bipartite systems remained elusive. The importance of just such states has been underlined recently by the discovery that they are a universal resource for measurement-based quantum computation~\cite{Wei}.

In this paper we study theoretically  multiorbital insulators described by Hubbard models with orbital-dependent directional nearest-neighbour hopping, and propose that they are a platform for realization of two dimensional VBS states.
The choice of honeycomb geometry is motivated by recent activity on the layered hexagonal Iridates such as Na$_2$IrO$_3$ and Li$_2$IrO$_3$. In these spin-orbit entangled Mott insulators the $t_{2g}$ electrons lead to \cite{PhysRevLett.105.027204} anisotropic spin-spin interactions of the  Kitaev-Heisenberg model. We consider a similar Hubbard model, however in a d$^3$ configuration, i.e. having three electrons per site (half filling); an example is Li$_2$MnO$_3$. While Li$_2$MnO$_3$ shows magnetic order at low temperatures \cite{Li2MnO3}, variants of this material may be in the VBS phase.

\begin{figure}
\centering
\includegraphics*[width=8cm]{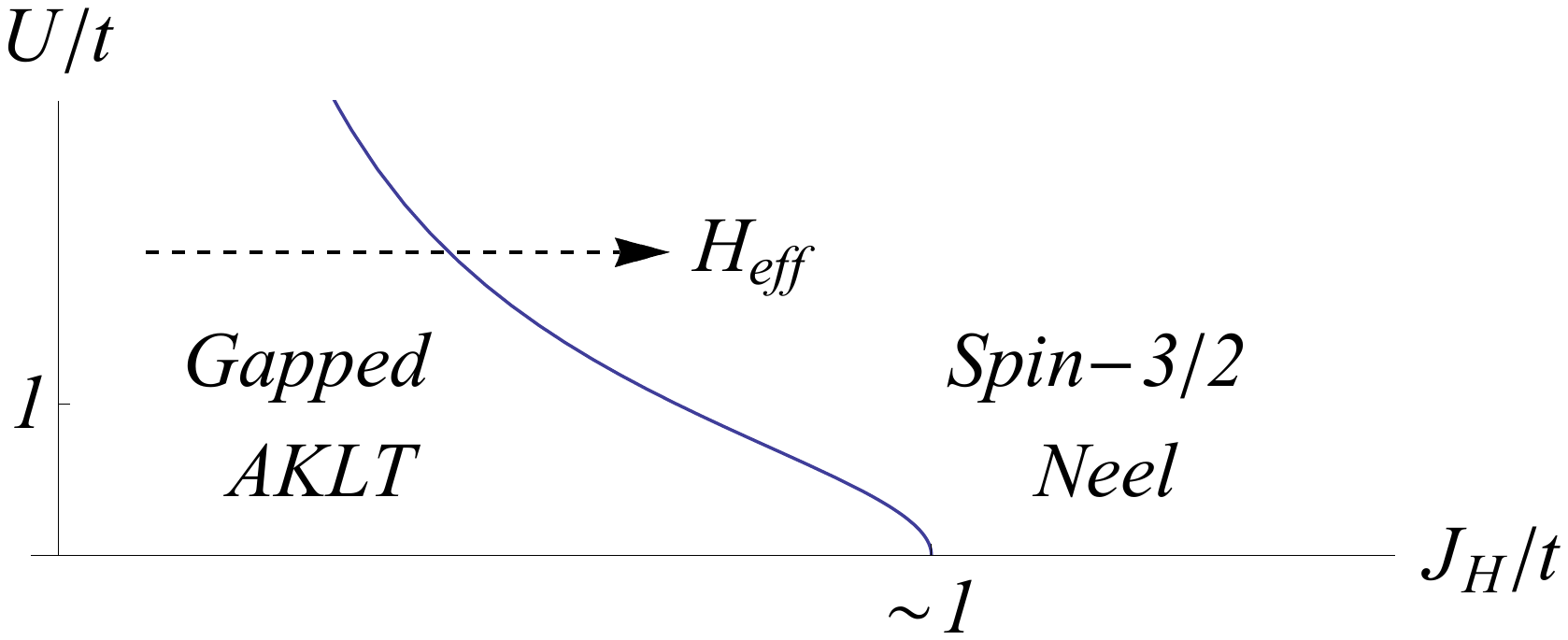}
\caption{ Phase diagram of the system Eq.(\ref{eq:model}) for $U' < U$.}
\label{fg:PDschematic}
\end{figure}

We show that the VBS state is realized in our model and is stable against interactions; it is adiabatically connected to the AKLT state. Using analytic arguments and DMRG simulations \cite{PhysRevLett.69.2863} we find that while strong Hund's rule coupling leads to 
bulk Neel order, an extended region in the phase diagram at small $J_H$ exists, with properties identical to the AKLT state (see Fig. \ref{fg:PDschematic}): an energy gap, exponentially decaying correlations, a singlet groundstate, and, notably, symmetry protected free spin-1/2 excitations at lattice vacancies, which provide an experimental signature.

\paragraph{Model \--}
We consider a tight-binding model of $t_{2g}$ electrons on a honeycomb lattice, arising for example in the basal plane of the Iridate compounds, as seen in Fig.~\ref{fig:t2g3plots}:
\begin{gather}\nonumber
H = -t \sum_{\sigma = \uparrow , \downarrow}\sum_{\langle i,j \rangle_\gamma} c^\dagger_{i,\gamma,\sigma} c_{j,\gamma,\sigma}+H.c. \mbox{\ \ \ \ \ \ \ \ \ \ \ \ \ \ \ \ }\\
\label{eq:model}
+ \sum_i \sum_{\gamma \sigma\ne\gamma' \sigma'} U_{\gamma \gamma'} n_{i ,\gamma, \sigma} n_{i ,\gamma', \sigma'} - J_H \sum_i \vec{s}_i^{\mbox{\ }2},
\end{gather}
where $\gamma=xy,yz,zx\equiv1,2,3$ is the orbital index, $n_{i,\gamma} = \sum_\sigma c^\dagger_{i,\gamma,\sigma} c_{i,\gamma,\sigma}$ and $\vec{s}_{i} = \sum_{\gamma} \vec{s}_{i,\gamma}$ with $\vec{s}_{i,\gamma} = \sum_{\sigma , \sigma'} c^\dagger_{i,\gamma,\sigma} \frac{\vec{\sigma}_{\sigma \sigma'}}{2}c_{i,\gamma,\sigma'}$. The notation $\langle i,j \rangle_\gamma$ implies that the hopping between nearest-neighbours $i$ and $j$ involves $\gamma$ orbital only.

The directional hopping originates from the anisotropy of the $t_{2g}$ orbitals, which are predominantly confined to the $xy$, $yz$, and $zx$ planes. Consequently there is an appreciable spatial overlap of wavefunctions on neighbouring sites along preferred directions only, as shown in Figs.~\ref{fig:t2g3plots}(abc), and the hopping matrix elements in nonequivalent directions on the honeycomb lattice are dominated by distinct $t_{2g}$ orbitals, see Fig.~\ref{fig:t2g3plots}(d).
For simplicity we neglect hopping elements other than those in Eq. (\ref{eq:model}), which we don't expect to change the picture.

The interaction terms contain a Hund's rule coupling $J_H$ favoring large spin formation, and a Hubbard interaction $U_{\gamma \gamma'} =U \delta_{\gamma \gamma'} +U'(1- \delta_{\gamma \gamma'})$. Here $U$ and $U'$ are the interactions between electrons residing in the same or distinct orbitals, respectively; we assume $U'<U$.
We focus on a configuration with $3$ electrons per site, as in Li$_2$MnO$_3$~\cite{Li2MnO3}.
\paragraph{Phase diagram \--}
Consider first the case of vanishing Hund's rule coupling $J_H=0$. 
In the noninteracting limit $U=U' \equiv 0$ the system trivially forms a dimer band-insulator (six electrons per unit cell); the ground state is a product of spin-singlets on links with an energy gap of $2t$. Allowing $U \not= 0$ while keeping $U'=0$ does not alter the picture, since the Hamiltonian only couples pairs of sites connected by a single orbital $\gamma$, i.e. it factorizes into a sum of decoupled dimer Hamiltonians $H_{\langle i,j \rangle_\gamma}$. One can show that the ground state remains a product of singlets for any value of $U/t$.

In the large $U$ limit
three spin-$1/2$ states are occupied at each site, in distinct orbitals, since $U' < U$.  The electron hopping generates an interaction between the spin degrees of freedom connected by the same orbital $\gamma$, as shown in Fig.\ref{fig:t2g3plots}(d): $H_{eff} = J \sum_{\langle i,j \rangle_\gamma} \vec{s}_{i,\gamma} \cdot \vec{s}_{j ,\gamma}$ with $J = \frac{t^2}{U}$. Thus at $J_H=0$, for any $U$, the ground state is a product of singlets with an energy gap $E_{gap} = 2 |t|$ for $t \gg U$ and $E_{gap}=\frac{t^2}{U}=J$ for $U \gg t$.
While one cannot solve the model analytically for intermediate $U$, we expect a gapped phase all along the line $J_H=0$, as illustrated in Fig.~\ref{fg:PDschematic}. 

Furthermore, at large $U/t$ and $J_H \ll U$ we can write down an effective  Hamiltonian for three spins-1/2 per site, as a function of $J_H$:
\be
\label{eq:toymodel}
H_{eff} =J \sum_{\langle i,j \rangle_\gamma} \vec{s}_{i \gamma} \cdot  \vec{s}_{j \gamma} -J_H \sum_j (\sum_\gamma \vec{s}_{i \gamma})^2.
\ee
In the limit of large $J_H/t$ a spin-3/2 degree of freedom is formed on each site $i$, denoted by $\vec{T}_i = \sum_{\gamma} \vec{s}_{i, \gamma}$.  They are coupled via a Heisenberg interaction $H =J_{{\rm{eff}}} \vec{T}_i \cdot \vec{T}_j$ with $J_{{\rm{eff}}} =\frac{J}{9}=\frac{t^2}{9U}$, since in the spin-3/2 subspace of the Hilbert space of three spins-1/2 we have $\vec{s}_{i\gamma} = \frac{1}{3} \vec{T}_i$.
The spins-$3/2$ live on a bipartite lattice and will thus order antiferromagnetically in this limit and break the spin SU(2) symmetry. Since at small $J_H/t$ the system is a gapped singlet, there is necessarily a phase transition as function of $J_H$. The transition should occur at $J_H \sim J = \frac{t^2}{U}$.

The spin-3/2 formation holds at large $J_H$ for any $U$. In the limit of $U \ll J_H$ their effective coupling is $J_{{\rm{eff}}} =   \frac{2 t^2}{7 J_H} $ (from $2^{nd}$ order perturbation theory).
 At $U=0$ the full model (\ref{eq:model}) contains only two parameters $(J_H,t)$, hence we expect that the phase transition will occur at $J_H \sim t$.
A schematic phase diagram is drawn in Fig.~\ref{fg:PDschematic}. We support it using DMRG simulations.

\begin{figure}
\centering
\includegraphics[width = 8.5cm]{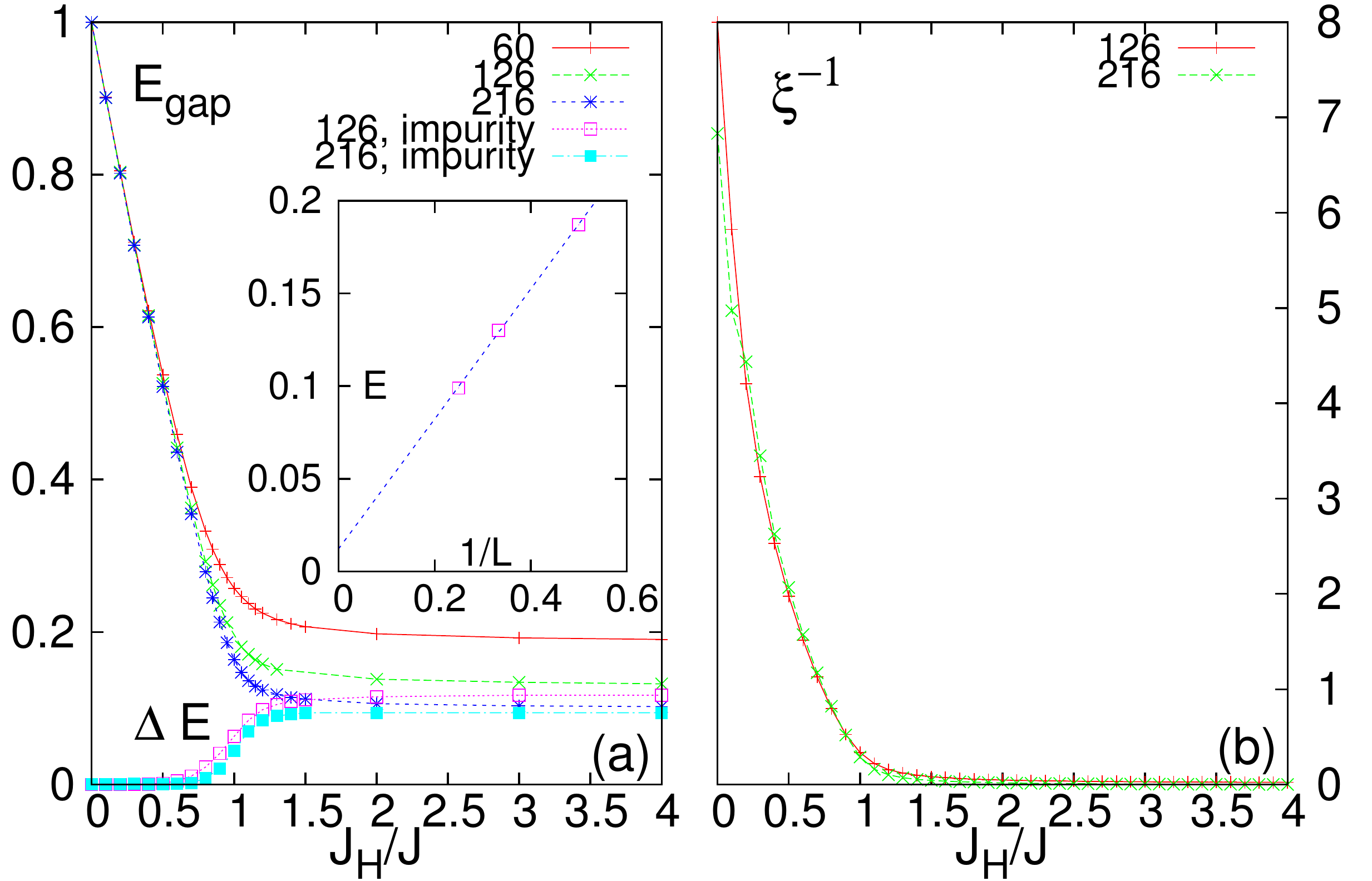}
\caption{(Color online) (a) Energy gap of model Eq.~(\ref{eq:toymodel}) in units of $J$ with and without an impurity. Inset: Finite size scaling of $E_{gap}$ for $J_H/J=10$, (b) Inverse correlation length $\xi^{{-1}}$ extracted from the exponential decay of $\langle s^z \rangle$ (see Fig. \ref{fig:DecEnt}a). }
\label{fig:EC}
\end{figure}
\paragraph{Numerical simulations \--} To study the phase diagram we performed density matrix renormalization group (finite DMRG) calculations using the ITensor (http://itensor.org) package. We simulated model Eq.~(\ref{eq:toymodel}) on a decorated honeycomb lattice using cylindrical boundary conditions and system sizes of up to 216 spins-1/2, keeping a fixed aspect ratio of the cylinder.

The calculated energy gap is shown in Fig. \ref{fig:EC}a: the system is gapped not only in the trivial case $J_H =0$, where the gap is equal to $\frac{t^2}{U}=J$, but also over a finite parameter range of $J_H < J$. Upon increasing $J_H$ the system reaches a transition at $J_H \sim J$. For finite size systems we simulated, the energy gap
saturates at a finite value in the Neel phase, corresponding to the finite size spectrum of spin wave excitations. It vanishes, however, in the thermodynamic limit, consistent with the infinite size extrapolation shown in the inset. At small $J_H$ the gap behaves as $E_{gap} = J - J_H+\mathcal{O}(J_H^2)$. This linear dependence can be derived analytically: consider an elementary excitation of the VBS state, which is a triplet on some link with an energy cost $E_{gap}=J$. In 1$^{st}$ order perturbation theory $J_H$ induces hopping of such excitations on the Kagome lattice formed by the links, hence reducing the energy gap as shown. The phase transition may be thus thought of as condensation of such spin-triplet excitations.

\begin{figure}
\centering

\includegraphics*[width=8.5cm]{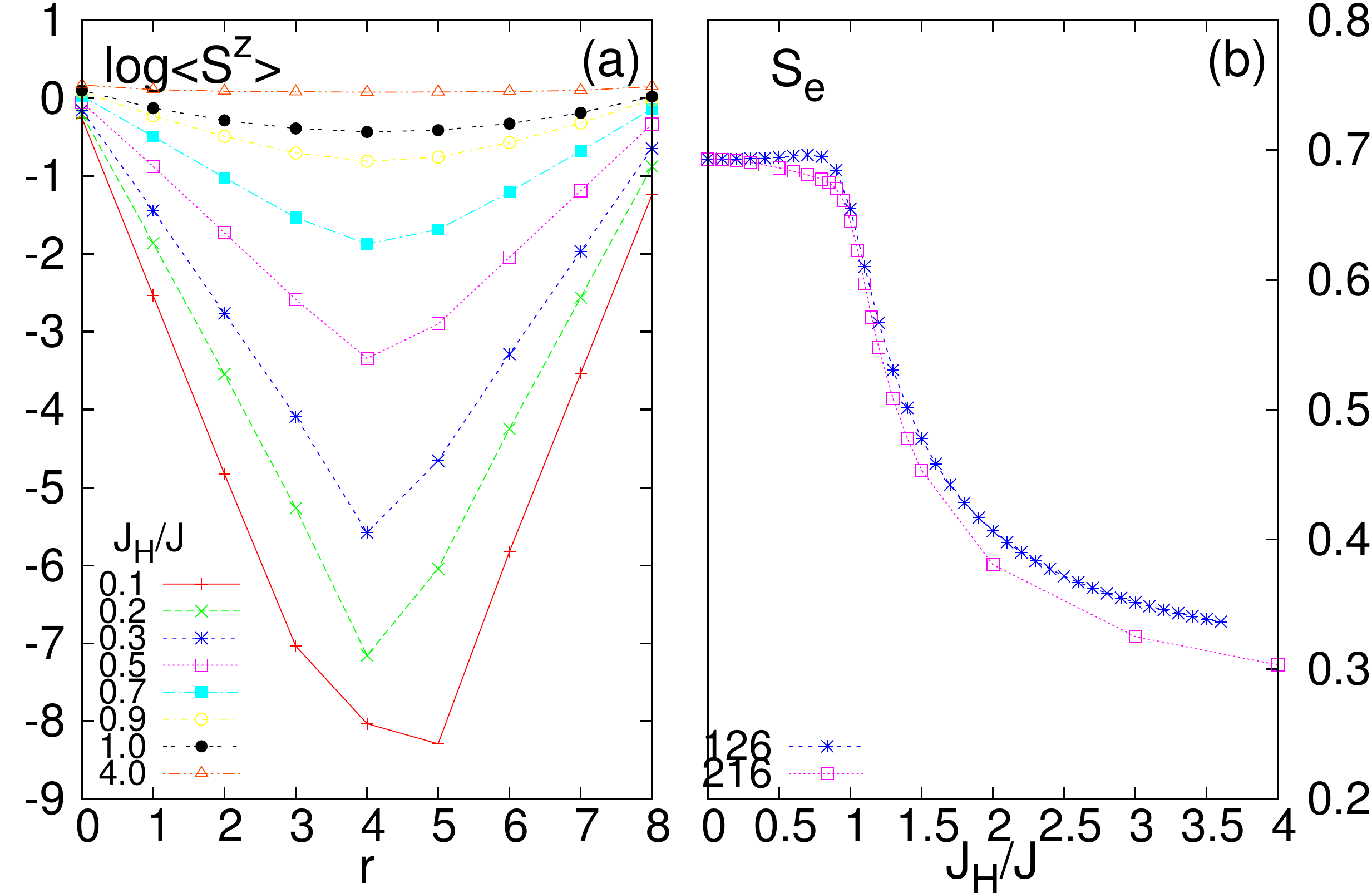}
\caption{(a) Expectation value $\sum_{\gamma} \langle s_{j,\gamma}^z \rangle$ on one of the sublattices, along the cylinder, 8 hexagons in length (b) Entanglement entropy $S_e$ of the ground state of model Eq.~(\ref{eq:toymodel}) normalized by the number of cut bonds.}
\label{fig:DecEnt}
\end{figure}

While in principle it is not obvious that there is a direct AKLT-Neel transition, the absence of an intermediate phase can be tested by comparing the gap closing with spin-spin correlations. Those can be numerically extracted by applying a strong magnetic field on both boundaries of the cylinder and measuring the AF order parameter $\langle s^z \rangle$ as function of distance from the boundaries. This is shown in Fig. \ref{fig:DecEnt}a for various values of $J_H$. For small $J_H$ there is exponential decay as expected for the AKLT state~\cite{PhysRevLett.59.799} while for large $J_H$ there is no decay as expected for the long range ordered Neel state. The extracted  inverse  AF spin-correlation length $\xi^{-1}$ is plotted in Fig. \ref{fig:EC}b; its vanishing coincides with the vanishing of the gap, ruling out the possibility of an intermediate phase. The behaviour of the entanglement entropy $S_e$ when partitioning the system into two halves provides another signature of a single phase transition: a singularity develops at $J_H=J$, as seen in Fig. \ref{fig:DecEnt}b.

\paragraph{Boundary excitations\--} One of defining properties of the AKLT state are the emergent degrees of freedom on system boundaries. Consider a lattice vacancy obtained by removing a single site from the bulk of the honeycomb lattice; see Fig.~\ref{fig:t2g3plots}(e).
In the AKLT state,  this breaks three valence bonds hence creating three unpaired spins-1/2 surrounding the vacancy. Those are generically coupled, however, as long as the system respects time reversal symmetry a Kramers-degenerate spin-1/2 degree of freedom must remain. This effect is reproduced in our numerical simulations of the model Eq. (\ref{eq:toymodel}) with a vacancy: we remove a site and compute the ground state energy in sectors whose total $S^z$ differs by one. The light blue and purple curves in Fig. \ref{fig:EC}a represent the energy difference $\Delta E$ of those two states: they are degenerate in the AKLT phase. In contrast, the two states split in the magnetically ordered Neel phase, where $\Delta E$ becomes energy to create a magnon excitation.

The free spin-1/2 degrees of freedom emerging at lattice vacancies in the AKLT phase can be detected experimentally due to their contribution to the free spin susceptibility, which will be proportional to the impurity density.

\begin{figure}[h]
\centering
\includegraphics*[width=8cm]{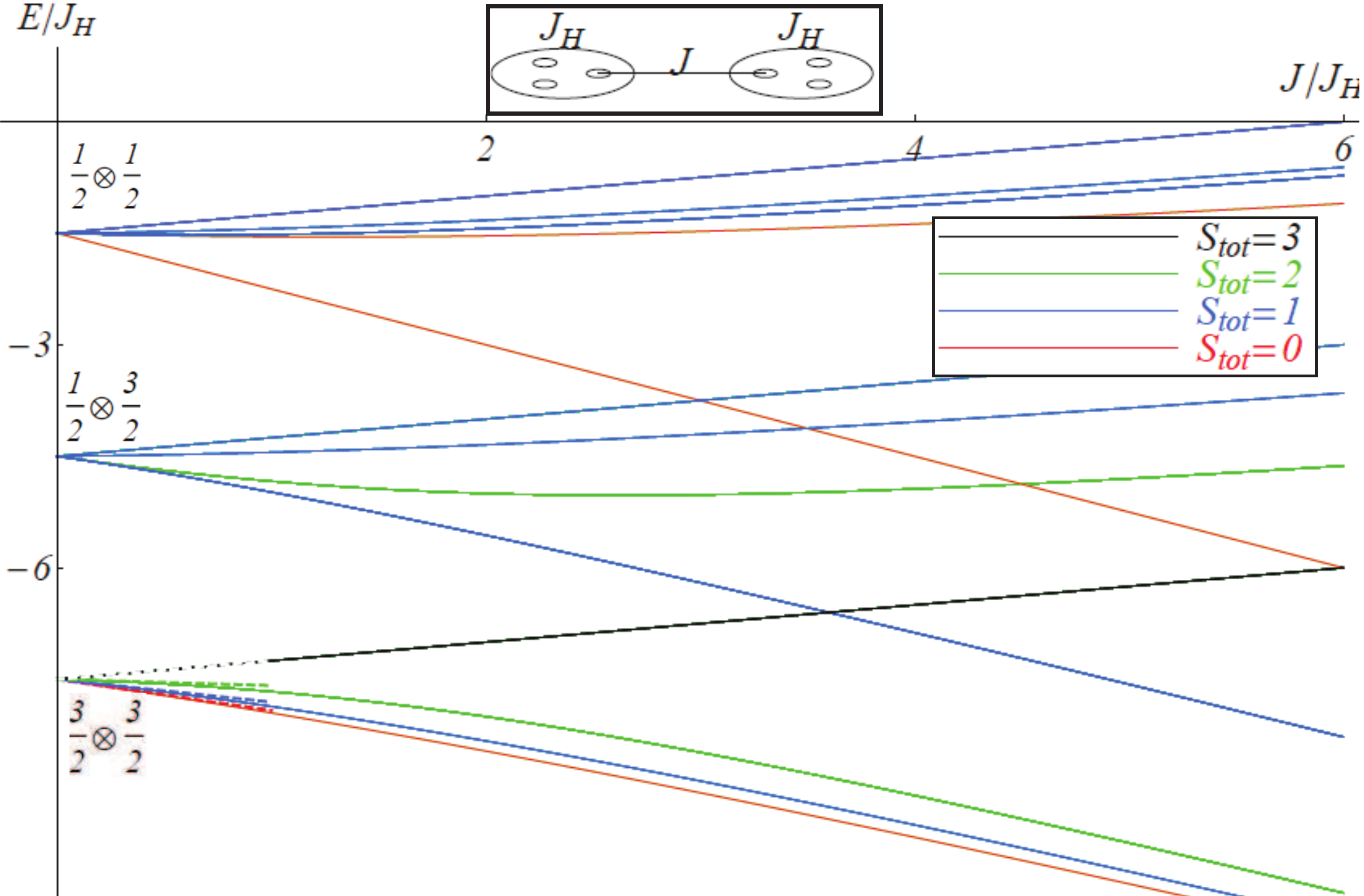}
\caption{Eigenvalues of the two-site Hamiltonian as a function of $J/J_H$. Inset above: two-site toy model.}
\label{fig:levelrepulsion}
\end{figure}

\paragraph{Two site toy model \--}
The mechanism by which interactions of the form of Eq.~(\ref{eq:AKLT32}) arise from the directional nature of the orbitals may be intuitively understood by considering a two-site toy version of the model Eq.~(\ref{eq:toymodel}), $H = - J_H (\sum_{\gamma} \vec{s}_{1,\gamma})^2 - J_H (\sum_{\gamma} \vec{s}_{2,\gamma})^2 + J \vec{s}_{1,1} \vec{s}_{2,1} $, see inset of Fig. \ref{fig:levelrepulsion}. 
There is only one $J$ coupling, which we take to be in the $\gamma=xy\equiv 1$ orbital.

The Hilbert space of three spins-$1/2$ at each site decomposes into two doublets and a quartet.
As we discussed before for large $J_H/J$ the spin-3/2 quartets are energetically favored by the Hund's rule coupling, and they are coupled by an effective Heisenberg AF interaction
$J_{{\rm{eff}}}=\frac{J}{9}$.
The Hilbert space of those two spins-3/2 decomposes in turn as follows: $\frac{3}{2} \otimes \frac{3}{2}  =  0 \oplus 1 \oplus 2 \oplus 3$, and consequently the Heisenberg term may be written in terms of projectors to subspaces of total spin:
 \be
  \vec{T}_1 \cdot  \vec{T}_2 = -\frac{15}{4} \mathcal{P}_0 - \frac{11}{4} \mathcal{P}_1 - \frac{3}{4} \mathcal{P}_2 + \frac{9}{4} \mathcal{P}_3.
 \ee
The energy splitting of those four multiplets is linear in J, as can be seen from dashed lines in bottom-left part of Fig. \ref{fig:levelrepulsion}, where the spectrum of this 6-spin model, computed via exact diagonalisation, is shown.
As expected, the ground state is a singlet. Note, however, that this result is valid for small $J/J_H$ only, when the doublets do not mix considerably with the spin-3/2 quartets.

Upon increasing $J/J_H$ the states belonging to doublets and quartets will mix; this can be seen in Fig. \ref{fig:levelrepulsion} in the form of level repulsion between the states of like total $S_{\rm{tot}}=0,1,2$ (see color code) originating in $\frac{3}{2} \otimes \frac{3}{2}$,$\frac{1}{2} \otimes \frac{3}{2}$ and $\frac{1}{2} \otimes \frac{1}{2}$ subspaces. Furthermore, the lowest $S_{\rm{tot}}=0,1,2$ levels are approximately degenerate -- they form a groundstate manifold, while the unique state of $S_{\rm{tot}}=3$ becomes an excited state. This mimics the spectral properties of the AKLT model Eq.~(\ref{eq:AKLT32}) which on two sites reads $H_{\rm{AKLT}} =\mathcal{J} \mathcal{P}_3$, in which all states but those forming a maximal total spin-3 between the two sites, are exactly degenerate.

The bunching of the lowest $S_{\rm{tot}}=0,1,2$ levels, a total of 9 states, can be intuitively understood in the limit of large $J/J_H$, when a singlet forms between the spins coupled by $J$. The remaining two pairs of spins-1/2 form two spins-1 due to the presence of $J_H$, whose Hilbert space is: $1 \otimes 1  =  0 \oplus 1 \oplus 2$. The approximate degeneracy between these $S_{\rm{tot}}=0,1,2$ states is broken once we include an isotropic, all orbital-to-all orbital coupling $J \sum_{ \gamma, \gamma'} \vec{s}_{1 \gamma} \cdot  \vec{s}_{2 \gamma'}$. 

\paragraph{Adiabatic connection of band insulator and the AKLT state \--}
We refer to the gapped phase in the diagram as the ``AKLT" phase since at $U = J_H=0$ the model can be smoothly connected in Hamiltonian space to the AKLT Hamiltonian, without closing the gap, as we now demonstrate. No other phase transitions intervening, the gapped phase of our diagram is therefore in the AKLT universality class.  The result is similar in spirit to Refs.~\cite{White96,PhysRevB.75.144420} connecting $S=1$ Heisenberg chain and the antiferromagnetic $S=1/2$ Heisenberg ladder, though in our intermediate transformations we break translation invariance.

At the noninteracitng point the ground state of our model Eq. (\ref{eq:model}) is a product of decoupled singlets. We show that the AKLT state can be connected to such a product of singlets.  The procedure works for AKLT spin models in arbitrary dimension. We work in the enlarged Hilbert space, where each AKLT spin-3/2 is composed of spin-1/2 degrees of freedom.
The decoupling into singlets is performed repeating the following steps:
(i) Choose a site $i$, and let $\langle ij \rangle$ for $j=1,2,3$ denote the three links incident on $i$. (ii) Adiabatically weaken the coupling constants $\mathcal{J}_{ij}$, until site $i$ is almost decoupled. The effective Hamiltonian  for site $i$ is given by spin-3/2 coupled to three effective spins-1/2 emerging at the boundary of the AKLT state around site $i$.
(iii) This Hamiltonian is adiabatically connected to three decoupled singlets, each made up of one of the effective boundary spins-1/2 and one of the three spins-1/2 which build the spin-3/2 on site $i$ (see supplemental material for a generic example of a path in Hamiltonian space).
(iv) Turn back the $\mathcal{J}_{ij}$ couplings. The three spins-1/2 at site $i$ are now  completely decoupled from each other (and only coupled to the spins-3/2 at neighbouring sites $j$). (v) Move to the next site.

Upon visiting all sites, the Hamiltonian will have become a sum of decoupled singlets on individual links, without ever closing the gap. We have thus managed to find a path in hamiltonian space between $H_{AKLT}$ and the noninteracting limit of our model Eq.~(\ref{eq:model}).

\paragraph{Conclusions and outlook \--}
We considered states which can be realized in electronic systems with orbital degrees of freedom, in which electron hopping, due to the directional character of the orbitals, is highly anisotropic. For a  half-filled 2D honeycomb lattice of $t_{2g}$ electrons we have shown that a 2D AKLT state may be realized as the ground state of the system. This state is known to be a universal resource for quantum computation \cite{Wei}.


The physics described in this paper, in particular the competition between the gapped singlet state of a valence bond solid (AKLT) type and that with a conventional long-range magnetic order, may be relevant for real correlated electron honeycomb systems. For example Na$_2$IrO$_3$~\cite{Choi12}  and Na$_2$RuO$_3$~\cite{Wang14} exhibit long-range magnetic order, but some samples of Li$_2$RuO$_3$ have singlet valence bonds~\cite{Miur07,Jackeli08}. Though this state is at a different filling to the one we consider, it raises the hope to realize the VBS state in similar compounds. We note that we neglected the effective dd hopping via ligand (e.g. oxygen p-orbitals), which may be relevant in some materials; also the  electron-lattice coupling may change the situation, leading to structural transitions with a loss of hexagonal symmetry, which seems to be the case in MoCl$_3$ \cite{ZAAC}.

The general picture described above (possibility of formation of AKLT-like state in a multi-orbital system) can be realized also in other geometries, beyond the 2D honeycomb lattice: it may apply to recently synthesized hyperhoneycomb lattices
 or to a new 3D structure of {$\gamma$}-Li$_2$IrO$_3$~\cite{Biffin}, and to novel hyperoctagon lattice structures proposed by M. Hermanns and S. Trebst~\cite{Hermanns}. In all these cases, for ions in $d^3$ configuration, we anticipate the possibility of existence of two competing states, the gapless state with a long-range magnetic order, and the gapful state of a VBS type.

\begin{acknowledgments}
We thank Erez Berg, Eli Eisenberg and Ady Stern for useful discussions. MKJ thanks E.M. Stoudenmire and Anna Kesselman for the helpful discussions on numerics.  This work was supported by ISF and Marie Curie CIG grants (ES).
\end{acknowledgments}


%
\end{document}


\title{Supplemental material: Two-dimensional Valence Bond Solid (AKLT) states from $t_{2g}$ electrons}

\maketitle

\paragraph{DMRG simulations \--}
We include here some technical details about the DMRG simulations, performed using the developer branch of the ITensor package (http://itensor.org) on the Weizmann Institute cluster. We have extended the functionality of the package by writing a new matrix-product Hamiltonian file for a decorated honeycomb lattice with and without a vacancy (see Figs. 1d,e in the main text). The computations were performed for cylindrical geometry, keeping the aspect ratio of cylinder length to circumference (in units of hexagons) equal to 2. The system sizes simulated were 60, 126 and 216 spin-1/2 degrees of freedom, which corresponds to $4$x$2$, $6$x$3$ and $8$x$4$ cylinders. The initial state for the DMRG algorithm was chosen to be a classical Neel state with $S^z_{tot}=0$. The free spin-1/2 degrees of freedom on the edges of the cylinder were frozen using a strong magnetic field applied there. These dangling free spins on the two edges of the cylinder live on two distinct sublattices of the Neel AF state, hence opposite magnetic fields were applied on different edges. We have kept up to $m=1600$ states in the $8$x$4$ computation and up to $m=800$ in the smaller ones.  The gap was obtained by comparing the ground state energies in sectors with $S^z_{tot}=0,1$ using the fact that the lowest excitation carries spin-1 for both the AKLT and the Neel phase.

\begin{figure}
\centering
\includegraphics*[width=7cm]{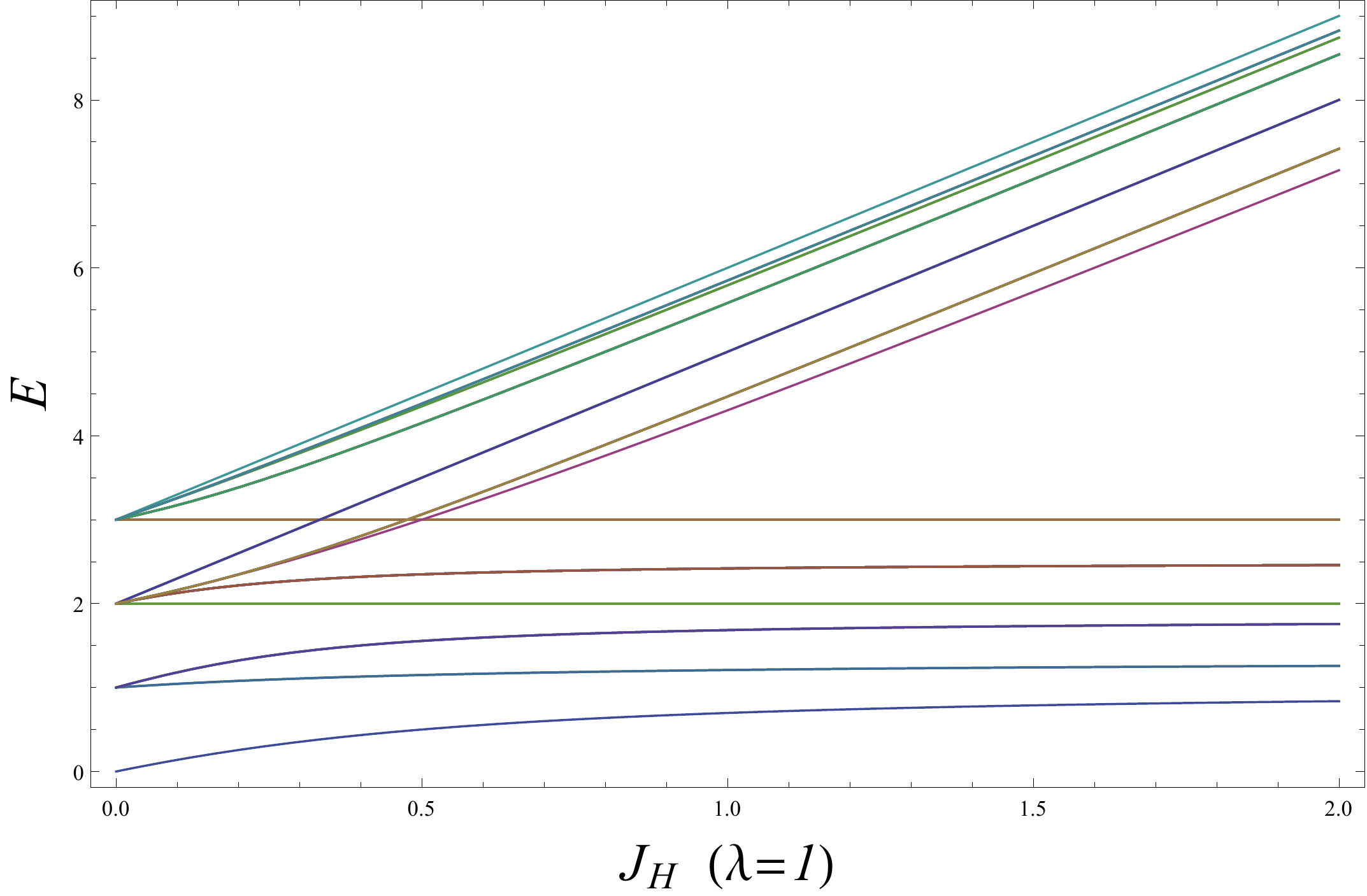}
\caption{Eigenvalues of $H[J_H,\lambda]$ in Eq.~(\ref{HJHLAMBDA}) as function of $J_H$ at $\lambda=1$, showing a singlet ground state for any $J_H$.}
\label{fig:3232}
\end{figure}

\paragraph{Adiabatic connection}

While it is not essential, for readers' interest, we show an explicit example of a path in Hamiltonian space mentioned in the main text, by means of exact diagonalization. The path itself is generic.


We write the central spin-3/2 as $\vec{S}=\sum_{\gamma=1}^3 \vec{\sigma}_\gamma$ and the three effective boundary spins we denote by $\vec{\tau}_\gamma$.
The projector into the total spin of $3/2+1/2$ between $S$ and one of the $\tau$ spins is:

\be
\mathcal{P}_{\frac{3}{2}+\frac{1}{2}}(\vec{S}, \vec{\tau}_\gamma) = \frac{1}{4} (\vec{S} + \vec{\tau}_\gamma)^2 - \frac{1}{2}.
\ee

Defining the interpolating Hamiltonian as:

\begin{gather}
H[J_H,\lambda]=-J_H\left( (\sum_{\gamma=1}^3 \vec{\sigma}_\gamma)^2 - \frac{3}{2}( \frac{3}{2}+1)\right) \nonumber \\
+ (1-\lambda) \sum_{\gamma=1}^3 \mathcal{P}_{\frac{3}{2}+\frac{1}{2}}(\vec{S}, \vec{\tau}_\gamma )
+ \lambda \sum_{\gamma=1}^3 (\vec{\sigma}_\gamma \cdot \vec{\tau}_\gamma+\frac{3}{4})\label{HJHLAMBDA},
\end{gather}
the initial Hamiltonian is $H_{\rm{init}} = H[\infty,0]$, and the final Hamiltonian is $H_{\rm{final}} = H[0,1]$. A path in the $(J_H,\lambda)$ parameter space should now connect these two points without closing the gap.

A convenient path is via $(J_H,\lambda)=(\infty,0) \to (\infty,1) \to (0,1)$. It is obvious that along the first transformation $(J_H,\lambda)=(\infty,0) \to (\infty,1)$ the ground state remains unchanged and the gap remains finite, since $\vec{S}$ is locked into a spin-3/2 state by $J_h$ and the Hamiltonian only undergoes a change by a constant and by a positive proportionality coefficient. The results of the second transformation $(\infty,1) \to (0,1)$ is shown in Fig.~\ref{fig:3232}, where the spectrum of the exactly diagonalized Hamiltonian is plotted as a function of $J_H$: the gap remains finite. We stress that this particular path was chosen for illustrative purposes only.